\documentclass[12pt]{article} \usepackage{amsmath, amssymb, amsfonts,
amstext, amsbsy, array, graphicx} \usepackage{float, epic, epsfig,
latexsym, curves, wrapfig, subfigure, graphics}

\newwrite\ffile\global\newcount\figno \global\figno=1

\def\writedef#1{}

\input epsf \def\figin{\epsfcheck\figin}\def\figins{\epsfcheck\figins}
\def\epsfcheck{\ifx\epsfbox\UnDeFiNeD \message{(NO epsf.tex, FIGURES
WILL BE IGNORED)}
\gdef\figin##1{\vskip2in}\gdef\figins##1{\hskip.5in}
instead \else\message{(FIGURES WILL BE INCLUDED)}%
\gdef\figin##1{##1}\gdef\figins##1{##1}\fi}

\def\figinsert{} \def\ifig#1#2#3{\xdef#1{fig.~\the\figno}
\writedef{#1\leftbracket fig.\noexpand~\the\figno}%
\figinsert\figin{\centerline{#3}}\medskip\centerline{\vbox{\baselineskip12pt
\advance\hsize by -1truein\center\footnotesize{ Fig.~\the\figno.} #2}}
\bigskip\endinsert\global\advance\figno by1}
\def\endinsert{}

\begin{document}
\baselineskip 18pt \newcommand{\Tr}{\mbox{Tr\,}}
\def\beq{\begin{equation}} \def\eeq{\end{equation}}
\newcommand{\bdm}{\[} \newcommand{\edm}{\]}
\newcommand{\bea}{\begin{eqnarray}}
\newcommand{\eea}[1]{\label{#1}\end{eqnarray}}
\renewcommand{\Re}{\mbox{Re}\,} \renewcommand{\Im}{\mbox{Im}\,}
\renewcommand{\a}{\alpha} \def\tm{\theta_-} \def\tp{\theta_+}
\def\fp{\phi_+} \def\fm{\phi_-} \def\dtq{(\partial_r\theta_-)^2}
\def\ptq{(\partial_\alpha\theta_-)^2} \def\dt{\partial_r\theta_-}
\def\pt{\partial_\alpha\theta_-} \def\N{{\cal N}}

\renewcommand\footnoterule{\vspace*{-3pt}
\hrule width 3in height 0.4pt \vspace*{5.6pt}}


\thispagestyle{empty}

\renewcommand{\thefootnote}{\fnsymbol{footnote}}

{\hfill \parbox{4cm}{ SHEP-05-29 \\ MPI-2005-110 }}

\bigskip

\begin{center} \noindent \Large \bf

Scalar effective potential for D7 brane probes which break chiral
symmetry

\end{center}

\bigskip\bigskip\bigskip

\centerline{ \normalsize \bf Riccardo Apreda $\negthinspace {}^a$, Johanna
Erdmenger $\negthinspace {}^a$ and Nick Evans $\negthinspace {}^b$
 \footnote[1]{\noindent \tt apreda@df.unipi.it, \\
 \hspace*{6.3mm}jke@mppmu.mpg.de, \\[1pt]
 \hspace*{6.3mm}evans@phys.soton.ac.uk .\\}}

\medskip\bigskip\bigskip

\centerline{ $^a$ {\it Max Planck-Institut f\"ur Physik (Werner
Heisenberg-Institut)} } \centerline{\it F\"ohringer Ring 6, D - 80805
M\"unchen, Germany}

\bigskip\bigskip

\centerline{ $^b$ {\it Department of Physics , Southampton
University}} \centerline{\it Southampton SO17 1BJ, United Kingdom}

\bigskip\bigskip\bigskip

\renewcommand{\thefootnote}{\arabic{footnote}}

\centerline{\bf \small Abstract}

\medskip

{\small \noindent We consider D7 brane probes embedded in deformed
  $AdS_5 \times S^5$ supergravity backgrounds which are
  non-supersymmetric in the interior. In the context of the
  generalised AdS/CFT correspondence, these setups are dual to
  QCD-like theories with fundamental matter which display chiral
  symmetry breaking by a quark condensate. Evaluating the D7 action
  for a surface instanton configuration gives
  rise to an effective potential for
  the scalar Higgs vev in the dual field theory. We calculate this
  potential for two specific supergravity backgrounds. For a metric
  due to Constable and Myers, we find that the potential is asymptotically
  bounded by a $1/Q^4$ behaviour and has a minimum at zero vev.  For the
  Yang-Mills$^*$ background we find that the Higgs potential scales quadratically
  with the Higgs vev. This corresponds to a canonical mass term and the
  embedding is again stable.}

\newpage


\addtolength{\skip\footins}{8mm}

\section{Introduction}

D7 brane probes have proved a versatile tool for including quark
fields into the AdS/CFT correspondence \cite{ads}. Strings
stretching between the D7s and the D3 branes of the original
AdS/CFT construction provide ${\cal N}=2$ fundamental
hypermultiplets \cite{add1}. Karch and Katz \cite{KK} proposed
that the open string sector on the world-volume of a probe D7
brane is holographically dual to quark--anti-quark bilinears $\bar
\psi \psi$. There have been many studies using probe D7s in a
variety of gravity backgrounds \cite{Myers1,d3d7}.  In this way a
number of non-supersymmetric geometries have been shown to induce
chiral symmetry breaking \cite{BEEGK,csb} (related analyses are
\cite{morecsb}), with the symmetry breaking geometrically
displayed by the D7 brane's bending to break an explicit symmetry
of the space. Meson spectra are also calculable \cite{spectra}.

In scenarios involving two or more D7 probes,
the Higgs branch spanned by squark vevs $\langle \bar q q
\rangle$ can be identified with instanton
configurations on the D7 world-volume \cite{Zach,Johannes}.
These configurations are the
standard four-dimensional instanton solutions living in the four
directions of the D7 world-volume transverse to the D3 branes.
The scalar Higgs vev in the field theory is identified with the
instanton size on the supergravity side
\footnote{Strictly speaking, it is a mixed Coulomb-Higgs branch which
  enters this duality, for details see \cite{Johannes}.}.
In the case of a probe in AdS space, there is a moduli space for
the magnitude of the instanton size or the scalar vev.  In less
supersymmetric gravity backgrounds though, the moduli space is
expected to be lifted. A potential may be generated, which either
may have a stable vacuum selecting a particular scalar vev, or may
have a  run-away behaviour. In a previous paper \cite{AEEG}, we
analyzed the Higgs branch of the ${\cal N}=4$ gauge theory at
finite temperature and density. For the finite temperature case we
found a stable minimum for the squark vev which undergoes a first
order phase transition as a function of the temperature (or
equivalently of the quark mass). On the other hand, in the
presence of a chemical potential the squark potential leads to an
instability, indicating Bose-Einstein condensation.  This implies
also for other supergravity backgrounds that the squark vev may
potentially be large, in which case physical observables such as
meson masses may be significantly modified.

In this paper we use the method developed in \cite{AEEG} to
analyze the scalar potential in the case of two quark flavours in
two gravity backgrounds describing gauge configurations that
induce chiral symmetry breaking. The first is the Constable-Myers
dilaton flow geometry \cite{Constable}(see also \cite{add2}). This
geometry was the first used to display chiral symmetry breaking in
\cite{BEEGK}, where it was used for its simplicity. The dilaton
flow retains an unspoilt S$^5$ which makes the probe analysis
particularly simple. From the field theory point of view, this
background has the essential property that it breaks supersymmetry
completely, such that the strong gauge dynamics may generate a
quark bilinear, which would be forbidden if some of the
supersymmetry were preserved.

The Constable-Myers geometry corresponds to the addition of a non-zero vev for
the operator $Tr F^2$ to the ${\cal N}=4$ gauge theory. This is expected to be unstable and therefore
the supergravity background is also expected to display
instabilities. Here however we show that the scalar quark
potential is well behaved and drives the vev to zero \footnote{Note
in the first preprint version of this paper we had mislaid a
factor of the dilaton which mistakenly suggested a run-away
potential.}. Asymptotically the potential has the form of a
constant minus a $1/Q^4$ term, with $Q^2 = \langle \bar{q} q \rangle$. This behaviour is determined
essentially by dimensional counting since the supersymmetry
breaking parameter, Tr$F^2$ is dimension four.

The Yang Mills$^*$ geometry \cite{Babington} is an example of a
more realistic dual. An equal mass term and/or condensate is
introduced for each of the four gauginos of the ${\cal N}=4$
theory breaking supersymmetry. Scalar masses are then generated
radiatively leaving a dual of pure Yang Mills theory in the
infra-red. The gaugino mass terms form an operator in the {\bf 10}
representation of SO(6), such that when the dual supergravity
scalar is switched on the S$^5$ of the geometry becomes crushed
(to two S$^2$). The embedding of a D7 brane in this geometry is a
complicated multi-coordinate problem and the full embedding is not
known. Nevertheless, in \cite{Waterson} it was shown that the core
of the geometry is repulsive to a D7 at all but one isolated point
in the parameter space of gaugino mass/condensate, which is the
expected signal for chiral symmetry breaking. Here we show that
for large Higgs vev, the effective potential for the squark vev is
quadratic and positive, indicating that small squark vevs are
favoured for the Yang-Mills$^*$ background. The UV of the theory
therefore has a standard scalar quark mass term. This is possible
because the supersymmetry breaking term, the gaugino mass is
dimension one.

\section{The Basic $N_f=2$ Squark Moduli Space}

Consider a probe of two coincident D7 branes in AdS${}_5\times S^5$.
This corresponds to two fundamental hypermultiplets in the dual gauge theory.
The metric of $AdS_5 \times S^5$ is given by
\begin{equation}
ds^2 = \frac{u^2}{R^2}\, dx_{//}^2 + \frac{R^2}{u^2} \, ( d \rho^2 +
\rho^2 d\Omega_3^2 + du_5^2 + du_6^2) \, ,
\label{adsmetric}
\end{equation}
where we have written the four coordinates on the D7 world-volume in
spherical coordinates ($\rho^2 = u_1^2 + u_2^2 + u_3^2 + u_4^2$).
$u_5$ and $u_6$ are the directions transverse to the D7 and
 \mbox{$u^2=\rho^2 + u_5^2 + u_6^2$}. $R$ is the AdS radius with $R^4 = 4 \pi g_s^2 N
\alpha^{'2}$. The dilaton and the four-form potential
read
\begin{equation}
e^{\Phi}=g_s, \qquad \quad C_{(4)}=  \frac{u^4}{g_s \, R^4} \, dx^0 \wedge
dx^1 \wedge dx^2 \wedge dx^3 .
\end{equation}
With $G_{ab}$ the pullback of the metric (\ref{adsmetric}),
the {\sl Einstein frame} action for the D7 probe is given by
\begin{equation}
S_{D7} = - \frac{T_7}{g_s^2} \int 
d^4 x \; d\rho\;
d \Omega_3\, \rho^3  e^{\Phi}\,   \Tr
\sqrt{-\det\left( G_{ab} + 2 \pi \alpha' e^{-\frac{\Phi}{2}} F_{ab}\right)} \, + \,  T_7 
\int 
C_{(4)} \wedge \Tr e^{2 \pi \alpha' F} .
\label{acD7ads}
\end{equation}
The regular solutions of the resulting
equations of motion for the embedding of the D7 are simply $u_5, u_6 = const.$
 (the quark mass is $m^2 = u_5^2 + u_6^2$).
\\
Next consider SU(2) instantonic solutions for the two-form $F_{ab}$ in (\ref{acD7ads}),
with the standard form \footnote{For our purposes it is enough to focus on the slice of the Higgs branch corresponding to a single, radially symmetric instanton.}
\begin{equation}
A_4 =\frac{i}{g} \frac{u_j \sigma_j}{Q^2+\rho^2},\quad \qquad A_j
=-\frac{i}{g} \frac{u_4 \sigma_j+\epsilon_{jkl} u_k
\sigma_l}{Q^2+\rho^2}, \qquad \quad j=1,2,3,
\label{instpot}
\end{equation}
where the $\sigma_i$ are the usual Pauli matrices, and we sum over
repeated indices. $Q$ is the instanton size, which is identified with the vev of the squark
fields in the dual gauge theory. For details see \cite{Johannes}.
The instantonic configuration
is self-dual with respect to a flat metric, and gives
\begin{equation}
\label{aci}
\Tr F_{mn} F_{mn} =-\frac{96}{g^2}\frac{Q^4}{(Q^2+\rho^2)^4} , 
\qquad m,n=u_1, \ldots , u_4.
\end{equation}
Expanding (\ref{acD7ads}) to second order in $F$ and dropping
the leading constant, we obtain
\begin{equation}
S_{D7}= -T_7 \; \frac{(2 \pi^2 \alpha')^2}{2\, g_s R^4}\int d^4x \; d
\rho \; \rho^3 (\rho^2+m^2)^2 \left( \Tr F_{ab} F_{ab} + \Tr F_{ab}^*
F_{ab} \right) \, .
\label{ai}
\end{equation}
Since for any instanton $F+F^*=0$, the two terms in (\ref{ai}) cancel exactly and we are left with the expected moduli space
of the ${\cal N}=2$ theory \footnote{Note there is a moduli space even in the
presence of a quark mass because the adjoint scalar vev can be set to
effectively remove that mass.}.

\section{Constable-Myers geometry}

The dilaton-flow geometry of
Constable and Myers \cite{Constable}
is asymptotically AdS at large
radius,  but is deformed in the interior of the space by an R-chargeless
parameter of dimension four ($b^4$ in what follows).  This geometry is
interpreted as being  dual to ${\cal N}=4$ gauge theory with a non-zero
expectation value for $\Tr F^2$. It  was used in
\cite{BEEGK}  to study
chiral symmetry breaking because of its particularly simple form with
a flat six-dimensional plane transverse to the D3 branes. The core of
the geometry is singular\footnote{This singularity may presumably be lifted by the D3 branes forming some sort of fuzzy sphere in the interior of the space.}.  D7
brane probes in the geometry are repelled by the central singularity
giving rise to chiral symmetry breaking.

\begin{figure}[H]
\begin{picture}(200,135)(0,0)
\put(2,110){\makebox(0,0)[b]{a)}}
\put(274,110){\makebox(0,0)[b]{b)}}
\put(230,0){\makebox(0,0)[b]{\footnotesize $\rho/b$}}
\put(30,123){\makebox(0,0)[b]{\footnotesize $u_5 (\rho/b)$}}
\put(476,0){\makebox(0,0)[b]{\scriptsize $Q/b$}}
\put(307,123){\makebox(0,0)[b]{\scriptsize $V(Q)/{\alpha'}^2$}}
\put(15,5){\epsfig{figure=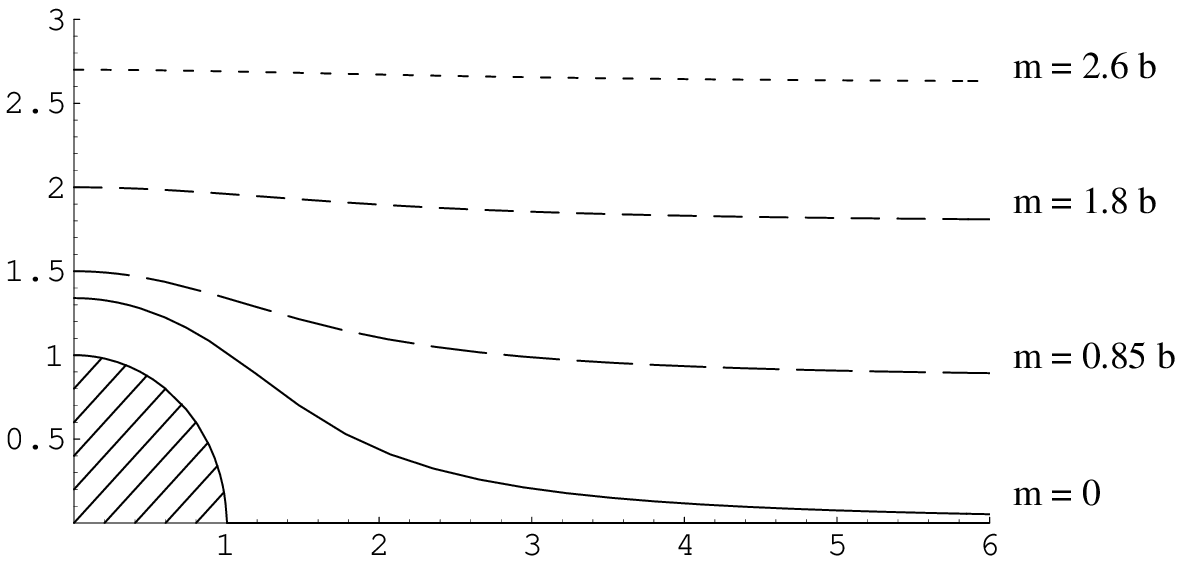,height=4cm}}
\put(286,5){\epsfig{figure=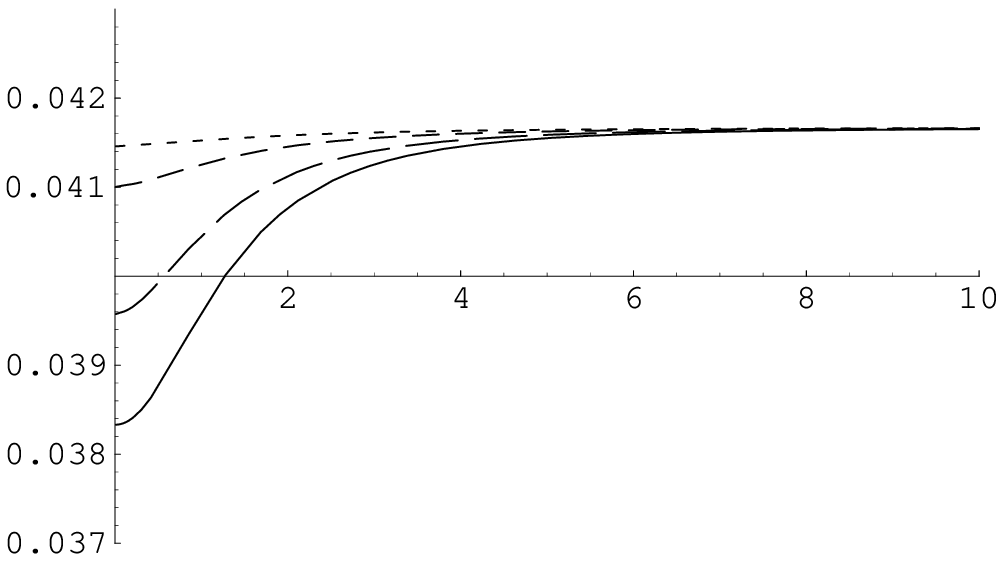,height=4cm}} \thinlines
\end{picture}
\caption{a) Plot of the D7 brane embedding in the Constable-Myers
 geometry for different values for $m_q$.
b) Potential versus  the Higgs vev $Q$ for the values of
$m_q$ shown in 1a).  All distances  are expressed in units of $b$, while  the potential is proportional to ${\alpha'}^2$.} \label{potincm}
\end{figure}

%
%
In {\it Einstein frame}, the Constable-Myers geometry is given by
\begin{equation}
ds^2 
=
H^{-1/2} 
K^{\delta/4}
dx_{4}^2 +
H^{1/2} 
K^{(2-\delta)/4} \frac{u^4 - b^4 }{ u^4 }
\sum_{i=1}^6 du_i^2,
\end{equation}
where
\[
K = \left( \frac{ u^4 + b^4 }{ u^4 - b^4}\right),
\qquad H = K^{\delta} - 1,
\qquad \quad \delta =\frac {R^{4}}{2 b^4}, \hspace{1cm} \Delta^2 = 10 - \delta^2 \, .
\]
$b$ is the deformation parameter.
 The dilaton and four-form are
\begin{equation}
e^{2 \Phi} = g_s^2 
K^{\Delta}, \hspace{1cm} C_{(4)} =
(g_s \,   H)^{-1} \,
 dt \wedge dx \wedge dy \wedge dz.
\end{equation}
We now place two D7 brane probes into the geometry,  such that
 they fill the $x_{4}$
directions and four of the $u$ directions (we write $\sum_{i=1}^4
u_i^2=\rho^2$).
The equation of motion that determines how the D7
lies in the $u_5$ direction as a function of $\rho$ is, with $u_6=0$,
\begin{equation}
\label{eommc} \frac{ d }{ d \rho} \left[ \frac{e^{\Phi} { \cal
G}(\rho,u_5) }{ \sqrt{ 1 + (\partial_\rho u_5)^2} } (\partial_\rho
u_5)\right] - \sqrt{ 1 + (\partial_\rho u_5)^2} \frac{ d }{ d
\bar{u_5}} \left[ e^{\Phi} { \cal G}(\rho,u_5) \right] = 0,
\end{equation}
\vspace{-3mm}
\begin{equation}
{\cal G}(\rho,u_5) = \rho^3 \frac{( (\rho^2 + u_5^{2})^2 + b^4) (
(\rho^2 + u_5^{2})^2 - b^4) }{ (\rho^2 + u_5^{2})^4}.\nonumber
\end{equation}
Asymptotically the solutions take the form $u_5 = m + c/ \rho^2 $.  The
regular solutions obtained numerically are shown in Figure
\ref{potincm}a). There is a U(1) symmetry in the $u_5-u_6$ plane which
corresponds to the axial symmetry of the quarks. For $m=0$ the
solution preserves this symmetry asymptotically, since
$u_5(\rho)$ goes to zero for large $\rho$.
On the other hand, the symmetry is broken in the interior,
where the solution $u_5$ is non-zero.
This is the signal of chiral symmetry breaking,
the D7 probe being repelled by the singularity in the interior.

Now we consider   the action of an instanton in the $u_1-u_4$
directions on the D7 world-volume with the above embeddings. In the
field theory this corresponds to computing the potential energy in the
non-supersymmetric theory with the scalar vev from the supersymmetric
theory's moduli space.
The induced metric on the embedded branes reads
\begin{equation}
G_{\mu\nu}= H^{-\frac{1}{2}} K^{\frac{\delta}{4}} \eta_{\mu\nu},  \qquad G_{u_i u_j}=  H^{\frac{1}{2}}K^{\frac{2-\delta}{4}}\frac{u^4-b^4}{u^4} \left(\delta_{ij}+ u_i u_j
\frac{
\left(
\partial_{\rho} u_5
\right)^2
}{\rho
^2
}
\right)\quad i,j= 1,\ldots ,4 \, ,
\label{inducedmetric}
\end{equation}
\vspace{-3mm}
\begin{displaymath}
\qquad \det G = K^2 \frac{(u^4-b^4)^4}{u^{16}} \left(1+(\partial_{\rho} u_5)^2\right)\, .
\end{displaymath}
We evaluate the D7-brane action on the space of fields strengths
which are self-dual with respect to the transverse part of the metric.
With the
coordinate transformation
\mbox{
$\tilde{u}^i=J(\rho) \, u^i, 
$
}
the induced metric
becomes a (conformally) flat metric 
provided that the function $J$ satisfies
\begin{equation}
\rho^2 \,(\partial_{\rho}J)^2 +2\,\rho \,J \,\partial_{\rho} J
- J^2\, \left(\partial_{\rho} u_5(\rho)\right)^2=0 \, ,
\end{equation}
together with the boundary condition $J(\rho\to\infty)=1$. In the
new coordinates $\tilde u^m$, the single instanton centered at the
origin has the usual form (\ref{instpot}), and the action is given by (\ref{aci}).

In the new set of coordinates we have, at second order in $\alpha'$,
\begin{equation}
S_{D7} = -T_7 \, \frac{ 6 (4\pi \alpha')^2}{g_s \, g^2} \int d^4 x \int d^4
  \tilde{u} \left[ 
\sqrt{-\det G } \; 
  G^{-2}_{\tilde{u}\tilde{u}}
   - \, g_s{} \, C_{(4)} \right]
 \frac{Q^4 }{ (\tilde{\rho}^2 + Q^2)^4 }
 \, .
\end{equation}
%
\noindent
In the original coordinates
\begin{displaymath}
S_{D7} = -T_7 \, \frac{6(4\pi \alpha')^2}{g_s \, g^2} \negthinspace  \int \negthinspace d^4 x \int \negthinspace d^4
  u  \,  \left[ 
\sqrt{-\det G } \; 
  G_{uu}^{-2} -  g_s{}^2 \, C_{(4)} \right]
  \frac{Q^4 J(\rho)^4 }{ (J(\rho)^2 \rho^2 + Q^2)^4 }\left(1+\frac{\rho\, J'(\rho)}{J(\rho)}\right)
\, .
\end{displaymath}
Since the instanton configuration is static,
the integral over the four $u$ coordinates  coincides with $-V$, the
 potential for the field $Q$. In polar coordinates we have
\begin{eqnarray}
 V(Q) &=& T_7 \, \frac{3(8\pi^2 \alpha')^2  } { g_s \, g^2} \int d \rho \;
  \rho^3 \left[\frac{Q^4 J(\rho)^4}{ ( J(\rho)^2\rho^2 + Q^2)^4 }\right]
\frac{Z^{\frac{\delta}{2}}-1}{Z^{\delta}-1}
\left( 1+ \rho\, \frac{J'(\rho)}{J(\rho)}\right)
\qquad  \label{potcmfin} \\[2mm]
Z&=&\frac{(\rho^2+u_5(\rho)^2)^2+b^4}{(\rho^2+u_5(\rho)^2)^2-b^4}\, .\nonumber
\end{eqnarray}

We plot this potential for different values of the quark mass
$m_q$ in figure \ref{potincm}b. For all $m_q$ the potential has a
stable minimum at zero vev. 

%
%
%
Both the IR and the UV behaviour of this solution may be checked analytically.  \\
Let us first consider the small $Q$ case.
It is always possible to split the $\rho$ integration of (\ref{potcmfin}) in the integration over the two intervals $\rho \le Q$ (or $\rho \in [0,Q]$ ) and $Q<\rho$ (or $\rho \in (Q,\infty) $).\\
At small $\rho$, both the embedding $u_5$ (see Fig. \ref{potincm}a ) and the function $J$ go to a constant plus corrections of order $\rho^2$, 
\[
u_5=u_5(0)+\frac{2 b^8- \Delta \, b^4 m^4}{2 m (m^8-b^8)}\, \rho^2+\ldots , \qquad \quad J=J(0)+ J(0) \frac{(2 b^8- \Delta\,  b^4 m^4)^2}{4 m^2 (m^8-b^8)^2} \, \rho^2+\ldots.
\]
Then we have for the first interval
\begin{equation}
V_{[0,Q]}\, =\, c_0
\int_0^Q \negthinspace  d \rho\; \rho^3 \; \frac{Q^4 J({\small 0})^4}{ (J(0)^2 \rho^2 +
Q^2)^4} \, (c_1+c_2 \; \rho^2+\ldots),
\end{equation}
where 
\[
c_0= \frac{ 3 T_7 (8\pi^2 \alpha')^2   }{g_s \, g^2} ,
\]
and, for example,
\[
c_1= \left(1+\left(\frac{u_5(0)^4+b^4}{u_5(0)^4-b^4}\right)^{\frac{\delta}{2}}\right)^{-1}.
\]
With the substitution $\tilde{\rho}=\rho/Q$ this integral is easily evaluated and gives
\[
V_{[0,Q]} = \frac{c_0 \, c_1}{24} +\frac{c_0 \, c_2}{48}\, Q^2+\ldots \ .
\]
In the second interval where $\rho > Q$ we may neglect the $Q$ contribution to the denominator of the integrand. For $Q$ small we have
\begin{equation}
V_{(Q,\infty)}\ =\  c_0\,
Q^4 \int
d \rho\; f(\rho) \,\, \sim \, \ c_0 \ \mathcal{I}\,
Q^4\, ,
\end{equation}
with $\mathcal{I}$ a numerical constant. 
Thus, summing up the two integration intervals, we recover the stable minimum for $V$ around $Q=0$.

The large $Q$ case is analogous.
Even without splitting the integration interval, 
 we may look at  large $Q$ solutions where the action will be dominated at large $\rho$. For large $Q$ and $\rho$ the action has the expansion
\begin{equation}
\label{mcint}
V =  \frac{ 3 T_7 (8\pi^2 \alpha')^2   }{g_s \, g^2} \negthinspace
\int \negthinspace \rho^3 d \rho\; \frac{Q^4}{ (\rho^2 +
Q^2)^4} \left[\; \frac{1}{2} - \frac {\delta}{4} \frac{ b^4 }{ u^4}+ \frac{1}{u^6}\frac{10 \, c^2+3 \; \delta\, b^4 \, m^2 }{ 6 }\; +\; \ldots \right].
\end{equation}
The asymptotic solution for the embedding has $u^2 = \rho^2 + u_5^2 =
\rho^2 + m^2+2m c/\rho^2+\ldots$.  Finally we rescale $\rho$ by Q to
make the integrals dimensionless. This gives
\begin{equation}
V =  \frac{ 3 T_7 (8\pi^2 \alpha')^2}{g_s \, g^2}
\left[\; \frac{1 }{ 2}\; {\cal I}_1 - \frac{b^4
}{ Q^4}\frac { \delta}{ 4 } \; {\cal I}_2 +
\frac{1}{Q^6}\frac{10\, c^2 + 3 \; \delta\, b^4 \, m^2 }{ 6} \; {\cal I}_3
 \, + \, \ldots \right] \, , \label{potasy}
\end{equation}
where the integrals, ${\cal I}_n$, are easily extracted from
(\ref{mcint}). 

It is interesting to use the same procedure of splitting the integration interval also in the large $Q$ case.
For $\rho \to \infty $ again both $u_5$ and $J$ go to a constant ($m$ and 1 respectively), such that 
\begin{equation}
 V_{(Q,\infty)} = \frac{c_0}{2} \int_Q^{\infty}  d \rho\; \rho^3\; \frac{Q^4 }{ (\rho^2 +
Q^2)^4}\; \left(1+O(\rho^{-2})\right) .
\end{equation}
With the substitution $\tilde{\rho}=\rho/Q$ we have
\begin{equation}
 V_{(Q,\infty)} = \frac{c_0}{2} \int_1^{\infty}  d \tilde{\rho}\; \frac{{\tilde{\rho}}^3}{ ({\tilde{\rho}}^2 +
1)^4}\; \left(1+O({\tilde{\rho}}^{-2})\right) \ = \  \frac{c_0}{48}+\ldots.
\end{equation}
Thus we see that the constant term of (\ref{potasy}) originates from the instanton configuration of size $Q$ probing the large $\rho$ region of the background, as it is natural since that region is asymptotically AdS where a flat potential as to be expected.
For the interval $[0,Q]$ we can instead neglect the $\rho^2$ term in the denominator of the integrand and we are left, for $Q$ large enough, with
 \begin{equation}
V_{[0,Q]}\ = \ c_0
 \int
  \negthinspace  d \rho\; \rho^3 \; \frac{ Q^4 J(\rho)^4}{ Q^8} \frac{Z^{\frac{\delta}{2}}-1}{Z^{\delta}-1}\ =
\ 
\frac{c_0}{Q^4} \int d \rho\; f(\rho) \, \ \sim \  \, c_0 \; \mathcal{I'}\, \frac{1}
{Q^4} \, .
\end{equation}
We see that the deviation from the flat potential of the AdS case originates from the instantonic configuration, even of very large size, probing the interior of the space.
Of course summing up the results for the two integration intervals we recover the first two terms of (\ref{potasy}).

We note that the expression (\ref{potasy}) vanishes for $b\rightarrow 0$ as
expected. The leading term is a constant independent of $Q$, which
corresponds to the fact that the metric returns to $AdS_5 \times
S^5$ for small $b$. This constant is also seen in the numerical
result shown in figure 1b). Moreover, the expansion (\ref{potasy})
successfully reproduces the form of the potential with a
$-b^4/Q^4$ term that forces the vev in towards zero.  In fact,
this behaviour is expected from dimensional analysis since the
deformation has dimension four and the non-constant part of the
potential must vanish as $b^4 \rightarrow 0$. This term with negative powers of the quantity associated to the squark bilinear operator indicates
that the field theory has a rather peculiar UV lagrangian with
self interactions involving inverses of the fields.
This is probably the result of integrating out the far UV completion, that is the $\mathcal{N}=4$ massive degrees of freedom.

We see that the screening effect which keeps the D7 embedding away from the 
singularity region also protects the open string sector modes associated 
to the squark vev from the instabilities which may arise  in the closed 
string sector due to the singular closed string background.

Let us provide some intuition for why the brane
configuration disfavours large instantons.
We suggest the
essential reason is that the background metric causes
volume elements to expand as $u=b$ is approached: The 
D7 brane bends away from the origin in order
to minimize its world-volume. The natural expectation is
that the instanton action for small instantons around
$\rho=0$ will grow with the size of the instanton preferring
zero size instantons. The same argument breaks down in pure
AdS because the four-form term conspires to cancel the
$\sqrt{\det G}$ volume term. When supersymmetry is broken, this
cancellation no longer works and the increase in the volume
term is the stronger effect.
This ensures a stable minimum for the Higgs vev.

\newpage

\section{Yang-Mills$^*$ geometry}

Next we consider the Yang-Mills$^*$ geometry \cite{Babington}.
This is a deformation of AdS by a supergravity field $\lambda$ in
the $\mathbf{10}$
of SO(6) and corresponds to an equal mass and/or
condensate for each of the four adjoint fermions of the ${\cal
N}=4$ gauge theory.
The adjoint scalar fields also acquire a mass.
This geometry
therefore naturally describes a pure Yang-Mills theory in the
infrared.

The geometry is complicated with the $S^5$ crushed to two $S^2$. This
implies that the D7 brane embedding has a complicated angular dependence, making
it hard to find the full D7 embedding. Therefore we restrict ourselves to
an asymptotic analysis, and check
that the scalar potential for the D7 probes is bounded.

\subsection{The Background}

The full 10d geometry reads \vspace{2mm}
\begin{equation}
ds_{10}^2 = (\xi_+ \xi_-)^{\frac{1 }{ 4} }ds_{1,4}^2+ (\xi_+ \xi_-)^{-\frac{3
}{ 4}} \, R^2 \, ds_{5}^2
\end{equation}
\[
ds_{1,4}^2 =e^{2 A(r)} \eta_{\mu\nu} dx^{\mu} dx^{\nu} + R^2 dr^2,
\hspace{1cm} ds_5^2 =\xi_- \cos^2 \alpha ~ d\Omega_{+}^2 + \xi_+
\sin^2 \alpha ~ d\Omega_{-}^2 +\xi_+ \xi_- d\alpha^2 
\]
with $d\Omega_{\pm}=d\theta_{\pm}^2+\sin^2\theta_{\pm}d\phi_{\pm}^2
$.
The $\xi_\pm$ are given by
\[
\xi_\pm = (\cosh \lambda )^2 \pm (\sinh \lambda)^2 \cos 2\alpha,
\]
\vspace{2mm}
%
%
%
%
The axion-dilaton field is purely complex so the $\theta$ angle is
zero
\begin{equation}
C + i e^{-\Phi} =\frac{i}{g_s} \sqrt{\frac{\xi_-}{\xi_+}}
\end{equation}
\noindent The two-form potential is given by
\begin{equation}
A_{(2)}=i R^2 \frac{\sinh 2 \lambda}{\xi_{+}} \cos^3 \alpha \, \sin
\theta_{+} d \theta_{+} \wedge d \phi_{+} - R^2 \frac{\sinh 2
\lambda}{\xi_{-}} \sin^3 \alpha \, \sin \theta_{-} d \theta_{-} \wedge
d \phi_{-}
\end{equation}
%
Finally the four-form potential lifts to
\begin{equation}
F_{(5)}= F +\star F, \hspace{0.5cm} F = \, dx^{0}\wedge dx^{1}\wedge
dx^{2}\wedge dx^{3}\wedge d\omega,
%
\hspace*{0.5cm} \omega(r)= \, g_s{}^{-1} \, R^4 e^{4A(r)} A'(r)\nonumber
\end{equation}
%
%
The fields $\lambda$ has a potential
$V=-\frac{3}{2}(1+\cosh^2\lambda)$ and, with $A$ satisfies

\begin{equation} \label{e1}
\lambda^{''} + 4 A^{'} \lambda^{'} = \frac{\partial V }{ \partial
\lambda},
\hspace{1cm} -3 A^{''} - 6 A^{'2} = \lambda^{'2} + 2 V
\end{equation}

\noindent
\mbox{Asymptotically, where the geometry returns to $AdS$, the solution we
are interested in behaves as}
\begin{eqnarray}
\lambda &=& m \, \frac{e^{-r}}{R}+\left(c+\frac{1}{6} \, m^3 ( 1+2 \, r)\right) \, \frac{e^{-3 r}}{R^3}+O(e^{-5r} R^{-5});
\nonumber\\ A &=& r-\frac{1}{6} \, m^2\, \frac{e^{-2 r}}{R^2} +
\left(
-\frac{1}{4}\, c\, m
-\frac{1}{18}\, m^4 %
\left(
1+\frac{3}{2} r%
\right) %
\right)
\,\frac{e^{-4 r}}{R^4}+O(e^{-6r} R^{-6})
\label{asy}
\end{eqnarray}
$m$ corresponds to an adjoint fermion mass term and $c$ to a
condensate. Numerically solving the equations of motion shows that
the geometry is singular in the interior, presumably reflecting
the presence of a fuzzy D3 configuration. However here we just
look at the UV behaviour where the geometry is dominated by the
fermion mass. This is straightforward using (\ref{asy}).

\subsection{Asymptotic Potential}

We now embed a probe of two coincident D7 branes into the Yang-Mills$^*$ geometry.
The full solution has dependence on $\alpha$
as well as on $r$, the latter being known only numerically. Therefore we restrict
to the asymptotic region of large $r$, where the embedding is known analytically.
Moreover we choose to place the D7 probe in the directions given by
$x_{//}$, $r$, $\alpha$ and $\Omega_+$. The choice of $\Omega_+$ rather than
$\Omega_-$ is motivated by the fact that   $\Omega_+$ supports the NS two-form, while
$\Omega_-$ supports the Ramond two-form. In this way we ensure that we consider
electrically instead of magnetically charged quarks.

For large $r$,  the
embedding is determined by
\begin{equation}
\sin \theta_-=M\frac{e^{-r}}{R \sin\alpha} +\ldots \, , \qquad \phi_-= 0 \, .
\label{emb}
\end{equation}
Here $M$ is the mass of the fundamental fermion
\footnote{In pure AdS we put the D7 along $u_5= R e^r \sin\alpha
\sin\theta_-=M$, \ \ $u_6=R e^r\sin\alpha \sin\theta_-\sin\phi_-=0$,
hence the  solution (\ref{emb}) in the new set of coordinates.}.

The full D7 action is
\begin{equation}\hspace*{-2mm}
S_{D7}^E = - \frac{T_7}{g_s^2} \int \negthinspace d^8\xi 
\; \Tr\negthinspace \left[e^{\Phi} \sqrt{-\det\left( 
G_{ab} + e^{-\frac{\Phi}{2}}
B_{ab}
+2 \pi \alpha' e^{-\frac{\Phi}{2}} F_{ab}\right)}
\right]+ \, T_7 \negthinspace \int \negthinspace C_{(4)} \wedge
\Tr \, 
e^{2\pi \alpha' F}
 \end{equation}
 with $d^8 \xi$ expressed in term of the embedding coordinates
$ d^4x \, dr\, d\alpha \, d \theta_+ d\phi_+$. $G_{ab}$ and $B_{ab}$
are the pullbacks of the metric and of the $B$ field respectively.

At lowest order in $\alpha' $ we find a term independent of the
instanton - and hence of the scalar vev $Q$ - which we drop. The linear
order in $\alpha'$ traces to zero, while at second order we have
\renewcommand{\arraystretch}{0.65}
\begin{eqnarray}
 S_{D7} &=& S_{DBI}+ S_{WZ} \, ,
\nonumber
\\[3mm]
S_{DBI}&=&  - T_7\, \frac{2 \, \pi^2 {\alpha'}^2}{g_s^2} 
\texttt{\Large{$\int$}} \negthinspace d^8 \xi\; \; \Bigg\{
 \sqrt{\frac{-\det G}{(1+ e^{-\Phi} B_{\tp\fp}B^{\tp\fp})(1+g_{\tm\tm}(\partial_r
     \theta_-  \partial^r \theta_- +\partial_\a \theta_-
\partial^\a \theta_-))}}
\qquad \qquad \qquad \qquad \qquad
\nonumber
\\[3mm]
&&
\Tr\bigg[
 \hspace{-7mm}
\sum_{\begin{array}{c} \text{ {\scriptsize $a\neq b$;}}\\
 \text{ {\scriptsize$ \ \  a,b \in (r,\a,\tp,\fp)$}}\end{array}}
 \hspace{-11mm}F_{ab}F^{ab}
 \;\; +\;\;\,
 g_{\tm\tm}
 \hspace{-12mm}
 \sum_{\begin{array}{c} \text{ {\scriptsize $a\neq b\neq c$;}}\\
  \text{ {\scriptsize $a\in(r,\a);$}}\\
  \text{ {\scriptsize $ b,c \in (r,\a,\tp,\fp)$}}\end{array}}
  \hspace{-11mm}\partial_a \theta_- \partial^a \theta_- F_{bc}F^{bc}
    \;\;- \; 2  \, g_{\tm\tm} \hspace{-4mm}
  \sum_{a\in (\tp,\fp)}
   \hspace{-3mm}
 \partial_r \theta_- \partial_{\a} \theta_- F^{ra} F^{\a}_{a} 
\nonumber
\\[1mm]
&& 
+ e^{-\Phi} B_{\tp\fp}B^{\tp\fp} \left(F_{r\a}F^{r\a} - F_{\tp\fp}F^{\tp\fp} \frac{(1+g_{\tm\tm}(\partial_r \theta_-  \partial^r \theta_- +\partial_{\a} \theta_- \partial^{\a} \theta_-))}{(1+ e^{-\Phi} B_{\tp\fp}B^{\tp\fp})}\right)\bigg]
\Bigg\} \, ,
\nonumber
 \\[3mm]
S_{WZ}&=& - T_7\, 2 \, \pi^2 {\alpha'}^2 \negthinspace
\texttt{\large{$\int$}} \negthinspace d^8 \xi\; \; C_{(4)} \Tr
\left[F_{ab}^* F_{ab}\right] \, .
\\[-4mm]
\nonumber
\end{eqnarray}
We now substitute  the asymptotics for the background
(\ref{asy}) and the embedding (\ref{emb}) in the previous formula
for the action. For the instanton field we use the standard
configuration (\ref{instpot}), expressed now in the $(r,\ \a,\
\tp,\  \fp)$ set of coordinates. Since we are only interested in
the asymptotic expansion, there is no need to introduce a
self-dual configuration with respect to the full induced metric.

After these substitutions the leading term in the action, without
$m$ dependence, is zero as in AdS due to opposing contributions
from the kinetic and the Wess-Zumino terms.

The first non zero contribution is second order in $m$:
\begin{equation}
S_{D7}= -  \, m^2\, \frac{ T_7 \, 2\, (4 \, \pi {\alpha'})^2}{3 \, g_s\,
g^2\, R^2} \int d^8\xi \; e^{6 r} \frac{R^4 Q^4}{(R^2 \, e^{2r}+Q^2)^4}\,
\cos^2\alpha \, (1+ 3 \cos 2\alpha)\sin\tp
\end{equation}
Thus, performing the
angular integrals and writing $R \, e^r/Q = w$, we get for the
potential
\begin{eqnarray}
V(Q) &=&\;\; m^2\, Q^2 \; \;\frac{5\, T_7 \, (8\, \pi^2 \alpha')^2}{6\,
g_s\, g^2 R^4} \int_{w_0}^{\infty} dw \,
\frac{w^5}{(w^2+1)^4} \nonumber \\[2mm]
 & \simeq &\;\; m^2\, Q^2 \; \;\frac{70\, T_7 \, (\pi^2
\alpha')^2}{ 9\, g_s\, g^2 \, R^4}
\end{eqnarray}
assuming in the last integral $ w_0 \simeq Q$.

The next correction to the potential is a constant term
proportional to  $m^2 M^2$. Further corrections are negligible,
being of the form of powers of $1/Q^2$ and even more suppressed by
additional factors of $m^2$. Though this result is derived at
large radius, and thus describes only the large $Q$ sector of the
field theory, the leading quadratic behaviour of the potential is
a good indication of the stability of the field-theoretical
vacuum.

\newpage

\noindent {\bf Acknowledgments:} We are very grateful to Zachary
Guralnik for collaboration at an early stage of this work. We also thank Christoph Sieg for very useful discussions on the D7 action. R.A.
acknowledges funding through the Deutsche Forschungsgemeinschaft
(DFG), grant ER301/2-1.

\end{document}